# Branched Broomrape Detection in Tomato Farms Using Satellite Imagery and Time-Series Analysis

Mohammadreza Narimani[a], Alireza Pourreza[*a], Ali Moghimi[a], Parastoo Farajpoor[a], Hamid Jafarbiglu[a], Mohsen Mesgaran[b]

[a] Department of Biological and Agricultural Engineering, University of California, Davis, CA, 95616, United States
[b] Department of Plant Sciences, University of California, Davis, CA, 95616, United States

## ABSTRACT

Branched broomrape (*Phelipanche ramosa* (L.) Pomel) is a chlorophyll-deficient parasitic plant that seriously threatens tomato production by extracting essential nutrients from its host, potentially reducing yields by up to 80%. Its primarily subterranean lifecycle and ability to produce over 200,000 seeds per plant—with seeds remaining viable in the soil for up to 20 years—make early detection essential for effective management. This study presents an end-to-end pipeline that utilizes Sentinel-2 satellite imagery and time-series analysis to identify broomrape-infested tomato fields in California. Regions of interest were defined using farmer-reported infestations, and satellite images with less than 10% cloud cover were selected. Twelve spectral bands were downloaded alongside sun-sensor geometry metadata, and 20 spectral vegetation indices (e.g., Normalized Difference Vegetation Index, Normalized Difference Moisture Index) were calculated. Additionally, five plant traits (e.g., Leaf Area Index, Canopy Chlorophyll Content) were derived using a neural network model calibrated with both ground truth and synthetic data. Canopy Chlorophyll Content trends were used to delineate transplanting-to-harvest periods, with phenological stages aligned using growing degree days derived from local weather data. Vegetation pixels were identified by masking out non-vegetative areas and then used to train a Long Short-Term Memory (LSTM) network on 18,874 pixels across 48 GDD time points. The model achieved 88% training and 87% test accuracy, with precision, recall, and F1 scores of 0.86, 0.92, and 0.89, respectively. Permutation feature importance analysis identified Normalized Difference Moisture Index, Canopy Chlorophyll Content, Fraction of Absorbed Photosynthetically Active Radiation, and Chlorophyll Red-Edge as the most informative features, consistent with the physiological effects of broomrape infestation on water and chlorophyll content. Our results demonstrate the potential of satellite-driven time-series modeling for detecting parasitic stress in tomato farms.

## 1. INTRODUCTION

Branched broomrape (*Phelipanche ramosa*) is a parasitic plant that poses a significant threat to tomato cultivation by attaching to the host's roots and siphoning essential nutrients, potentially leading to yield reductions of up to 80% [1]. The parasite's life cycle is predominantly subterranean, making early detection challenging. Furthermore, each broomrape plant can produce up to 200,000 seeds that remain viable in the soil for up to 20 years, complicating management efforts [2].

Traditional control methods, such as chemical herbicides, have been employed to manage broomrape infestations. However, these approaches often involve uniform applications that can be economically inefficient and environmentally detrimental. Moreover, the effectiveness of chemical control is limited due to the parasite's subterranean nature and the difficulty in timing applications to target the parasite effectively [3].

Recent advancements in remote sensing technologies offer promising avenues for detecting and managing broomrape infestations. Studies have demonstrated the utility of unmanned aerial vehicle (UAV) multispectral imaging to detect broomrape-infected sunflowers by analyzing temporal patterns in spectral vegetation indices [4]. Similarly, hyperspectral imaging has been employed to facilitate early detection of Orobanche cumana parasitism on sunflowers under field conditions [5]. These approaches leverage the subtle changes in canopy reflectance associated with broomrape-induced stress, enabling earlier intervention than traditional methods.

In tomato cultivation, drone-based multispectral imaging combined with deep learning techniques has been explored for the timely detection of branched broomrape [6]. This method integrates multispectral imagery with Long Short-Term Memory (LSTM) networks to analyze sequential growing stages, achieving notable accuracy in identifying broomrape-infested plants. While these UAV-based methods show promise, their scalability is limited due to the logistical challenges and costs associated with drone operations over large agricultural areas.

Satellite-based remote sensing, mainly through Sentinel-2 imagery, offers a scalable and efficient approach for monitoring large agricultural areas [7]. Sentinel-2 provides high-resolution multispectral data with a 5-day revisit frequency, making it well-suited for agricultural monitoring. By analyzing time-series data from Sentinel-2, it is possible to detect changes in spectral vegetation indices and plant traits indicative of broomrape infestation. Building upon these advancements, our study aims to develop an automated pipeline for detecting broomrape infestation in tomato farms using Sentinel-2 satellite imagery and time-series analysis. The specific objectives are to create a robust system for downloading region-specific Sentinel-2 imagery—including 12 spectral bands and key metadata (sun and sensor zenith/azimuth angles)—for defined tomato farm regions in California, filtering out images with more than 10% cloud cover; integrate the computation of 20 spectral vegetation indices and derive five neural network–based traits (Leaf Area Index [LAI], Leaf Chlorophyll Content [$C_{ab}$], Canopy Chlorophyll Content [CCC], Fraction of Absorbed Photosynthetically Active Radiation [FAPAR], and Fractional Vegetation Cover [FCOVER]) to accurately capture critical indicators of vegetation health and detect broomrape-induced stress; and utilize a Long Short-Term Memory (LSTM) model to analyze vegetation pixel data across 48 growing degree day (GDD) stages, aligning phenological phases and validating performance through cross-validation and permutation feature importance analysis, to achieve early detection of broomrape infestation. By achieving these objectives, we aim to enhance the precision and timeliness of broomrape detection in tomato farms, facilitating more effective management strategies and mitigating potential yield losses.

## 2. METHODOLOGY

### 2.1. INTEGRATED SATELLITE DATA PROCESSING AND FEATURE EXTRACTION FOR PHENOLOGICAL ANALYSIS

Five broomrape infested (identified and reported by farmers) tomato fields were compared to five non-infested fields in the same region. For each field, Sentinel imagery across the tomato growing season were monitored, capturing key phenological stages essential for early detection. A cloud cover threshold (<10%) was applied to retain only clear-sky images. The 12 spectral bands—from visible to shortwave infrared—were used to compute spectral vegetation indices. In addition, sun and sensor zenith/azimuth angles, along with the spectral bands, were provided as inputs to the pretrained neural network model. All data are exported as GeoTIFFs for integration into downstream analysis. From this imagery, two complementary feature sets were derived: (1) 20 widely used spectral vegetation indices, such as Normalized Difference Vegetation Index (NDVI) and Enhanced Vegetation Index (EVI), to monitor plant vigor, stress, and water status, and (2) five pretrained neural network–derived plant traits—Leaf Area Index (LAI), Leaf Chlorophyll Content ($C_{ab}$), Canopy Chlorophyll Content (CCC), Fraction of Absorbed Photosynthetically Active Radiation (FAPAR), and Fractional Vegetation Cover (FCOVER)—estimated using spectral and sun and sensor zenith/azimuth angles input [8]. For a complete list of vegetation indices and plant traits used in this study, please refer to Appendix Tables A1 and A2.

Critical growth stages were determined by tracking seasonal variation in Canopy Chlorophyll Content (CCC). For all fields, ground-truth transplanting and harvest dates were available and aligned well with characteristic patterns in the CCC time series—specifically, a steep increase in CCC following transplanting, a gradual rise to a maximum during peak vegetation, and a sharp decline near harvest. The maximum CCC value was used to indicate the peak vegetation stage, corresponding to the densest canopy cover. This temporal information enabled alignment of CCC-derived curves with reported field events, allowing for phenological synchronization across fields. By anchoring each field to its transplanting, peak, and harvest stages, we established a consistent framework for comparing fields at equivalent points in the crop growth cycle. Figure 1 illustrates the seasonal CCC trajectory alongside representative imagery from the transplanting, peak vegetation, and harvest phases.

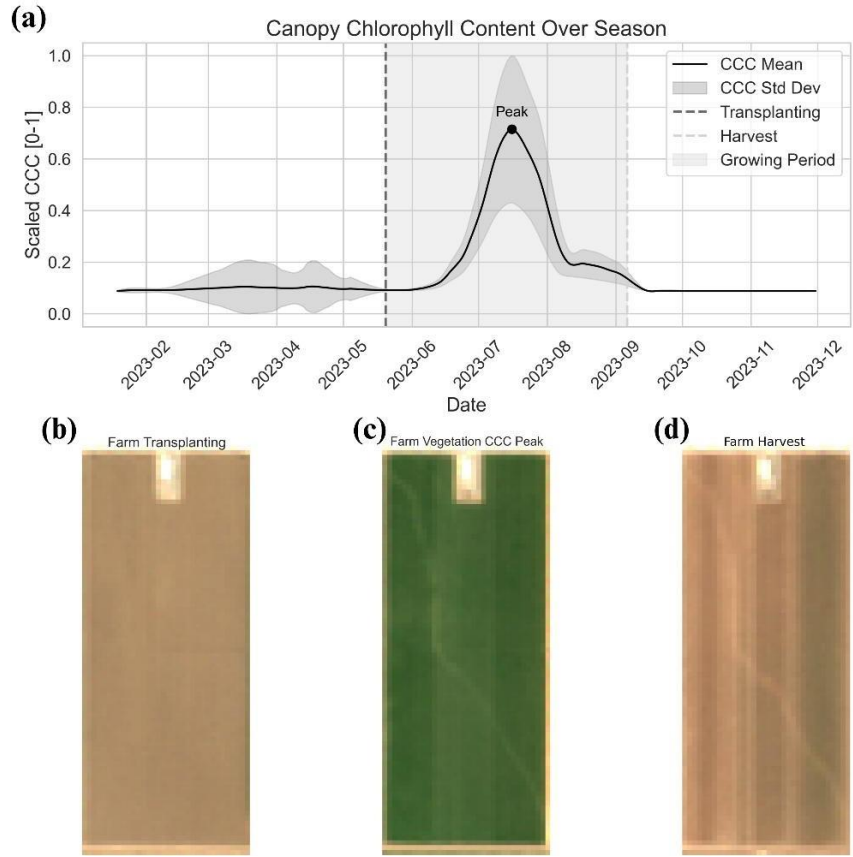

Figure 1: (a) Seasonal trend of scaled Canopy Chlorophyll Content (CCC) showing the transplanting, peak vegetation, and harvest phases. The shaded grey region indicates the growing period. (b)–(d) Corresponding satellite imagery of the farm during (b) transplanting, (c) peak vegetation CCC, and (d) harvest.

## 2.2. GDD-BASED PHENOLOGICAL ALIGNMENT, VEGETATION SEGMENTATION, AND INFESTATION MODELING

To standardize comparisons across tomato farms with varying transplanting and harvest dates, we employed Growing Degree Days (GDD) to represent accumulated thermal time, allowing alignment of phenological stages irrespective of calendar variations or local climate conditions. GDD was calculated by averaging the daily maximum and minimum temperatures and subtracting a base temperature critical for tomato development [9]. Daily temperature profiles were obtained from the Open-Meteo API (https://open-meteo.com/), and cumulative GDD was computed for each field from transplanting to harvest. Following alignment, we segmented vegetation from non-crop elements using five neural networks–derived plant traits—LAI, $C_{ab}$, CCC, FAPAR, and FCOVER—captured at the peak vegetation stage. Principal Component Analysis (PCA) was applied to reduce dimensionality, and K-means clustering was then used to classify pixels as either vegetation or background (e.g., soil or infrastructure). This approach effectively filtered out irrelevant signals, ensuring that only biologically meaningful crop pixels were passed to the model for infestation prediction.

After isolating vegetation pixels and compiling a comprehensive feature set—including 12 Sentinel-2 bands, 20 spectral vegetation indices, and five neural network–derived traits—across 48 GDD-based time steps, we employed a Long Short-Term Memory (LSTM) network [10] to classify each pixel as infected or non-infected. The model architecture consists of two LSTM layers with 64 and 32 units, interleaved with dropout layers for regularization, followed by fully connected layers leading to a binary classification output totaling 39,617 trainable parameters. The network was trained using five-fold cross-validation with a data split of 65% training, 15% validation, and 30% testing over 100 epochs. The LSTM effectively identified subtle physiological changes associated with broomrape infestation by capturing temporal dependencies in the canopy signals, enabling sensitive and early detection. The model was implemented on Google Colab using a Tesla T4 GPU with CUDA Version 12.4 and 15,360 MiB of memory, providing sufficient computational power for efficient training and evaluation. Figure 2 illustrates the architecture of the LSTM model and its sequential processing of the input feature stack.

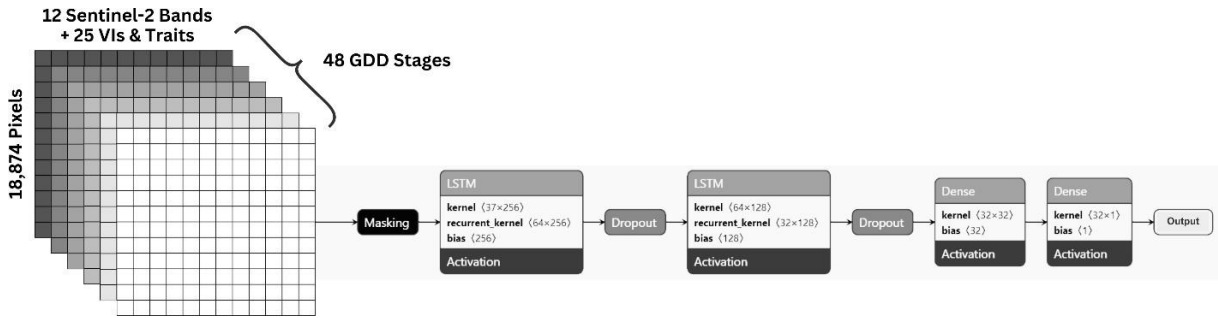

Figure 2: The overview of the LSTM architecture illustrates the input feature stack (37 features over 48 GDD stages) and sequential layers leading to the final classification output.

## 3. RESULTS AND DISCUSSION

The LSTM model was trained for 100 epochs using five-fold cross-validation, with the dataset split into 65% for training, 15% for validation, and 30% for testing. As shown in Figures 3a and 3b, both training and validation accuracy steadily increased, reaching a final training accuracy of 88%, while the corresponding loss curves declined consistently—indicating effective learning without overfitting. On the held-out test set, the model achieved an accuracy of 87%, precision of 0.86, recall of 0.92, and an F1 score of 0.89 (Figure 3d). The normalized confusion matrix (Figure 3c) shows strong agreement between predicted and true labels, particularly for the non-infected class, while the high recall emphasizes the model's ability to detect infected pixels reliably—critical for early broomrape detection. These results confirm the robustness of the LSTM model in identifying infestation patterns, supporting its application for timely intervention and improved tomato crop management.

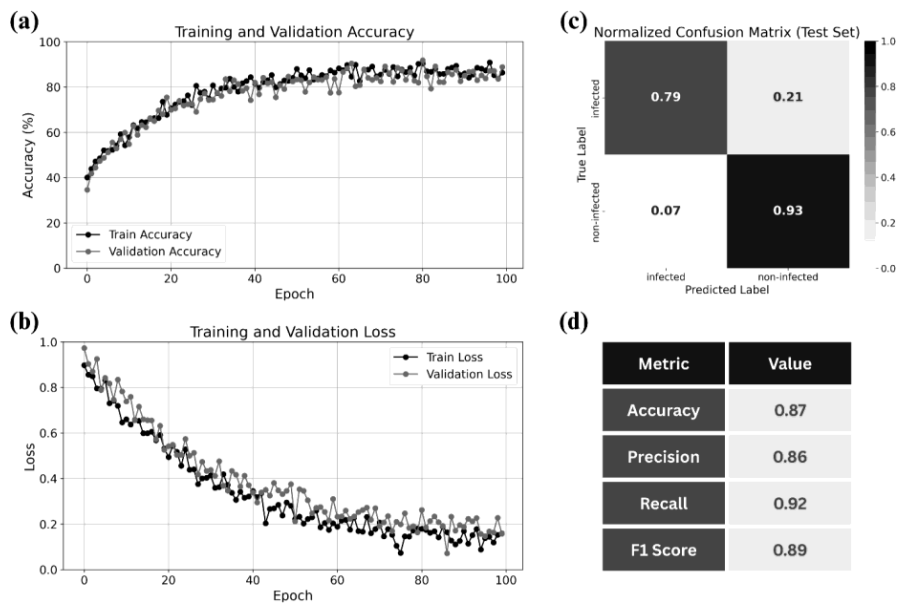

Figure 3: (a) Training and validation accuracy; (b) training and validation loss; (c) normalized confusion matrix on the test set; and (d) classification performance metrics—accuracy, precision, recall, and F1 score—summarizing the performance of the LSTM model.

In addition, a permutation feature importance technique [11] was applied directly to the LSTM model to assess the contribution of each input feature to the overall classification performance. By systematically permuting individual features and measuring the resulting drop in accuracy, we quantified the influence of each feature on the model's predictions. Four features emerged as particularly critical—NDMI (Normalized Difference Moisture Index), CCC (Canopy Chlorophyll Content), FAPAR (Fraction of Absorbed Photosynthetically Active Radiation), and CHL-RED-EDGE—which align with the biological effects of broomrape infestation. The parasitic plant reduces water availability and disrupts chlorophyll production, resulting in lower NDMI, CCC, and CHL-RED-EDGE, while FAPAR serves as an indicator of overall canopy vitality and photosynthetic efficiency. Histograms further confirmed these findings by consistently showing higher values for these indicators in healthy plants compared to infested ones; Figure 4 illustrates the permutation feature importance results.

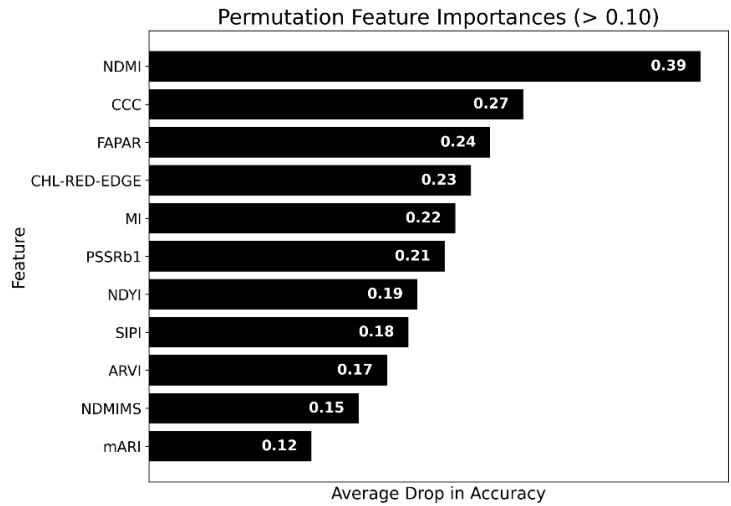

Figure 4: Permutation feature importance results, highlighting NDMI, CCC, FAPAR, and CHL-RED-EDGE as the most influential features in detecting broomrape infestation.

Key feature distributions between infected and non-infected fields were compared at the peak vegetation stage by analyzing NDMI, CCC, FAPAR, and CHL-RED-EDGE. Kernel density plots indicate that non-infected farms consistently exhibit higher values for all four features, reflecting more substantial water status (NDMI), healthier chlorophyll levels (CCC and CHL-RED-EDGE), and more robust canopy coverage (FAPAR). These observations support the biological premise that broomrape parasitism reduces water and nutrient availability [12], resulting in measurable declines in these key indicators, as illustrated in Figure 5.

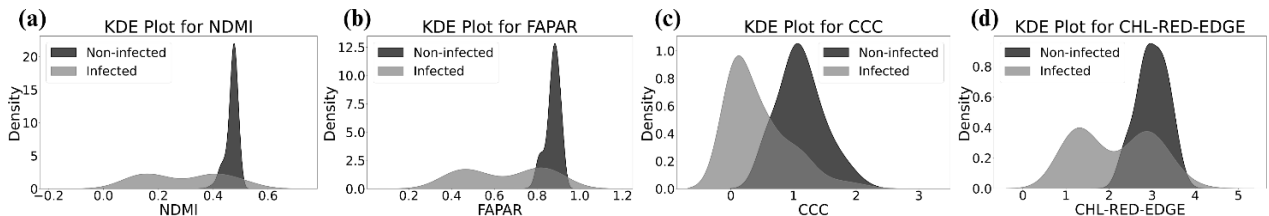

Figure 5: Kernel density plots comparing infected and non-infected farms at the peak vegetation stage for (a) NDMI, (b) FAPAR, (c) CCC, and (d) CHL-RED-EDGE, with non-infected farms showing higher overall values.

## 4. CONCLUSION

This study demonstrates a scalable approach for detection of branched broomrape infestation in tomato farms using Sentinel-2 satellite imagery and time-series analysis. A LSTM model, trained on time-series spectral feature, achieved promising performance (88% training accuracy and 87% test accuracy, with precision, recall, and F1 scores of 0.86, 0.92, and 0.89, respectively.

Key spectral features for brooomrape detection identified by a permutation feature importance include NDMI, CCC, FAPAR, and CHL-RED-EDGE, that reflects the known biological impacts of broomrape on water and chlorophyll content. Kernel density analyses also validated these findings by demonstrating consistently higher values for these features in healthy fields. Overall, this research highlights the potential of satellite-driven time-series modeling for early detection of emerging stresses that could facilitate timely intervention and treatment.

# 6. APPENDIX

Appendix Table A1. Spectral Vegetation Indices Used in This Study

| Index Acronym | Full Name |
| --- | --- |
| NDVI | Normalized Difference Vegetation Index |
| ARI | Anthocyanin Reflectance Index |
| mARI | Modified Anthocyanin Reflectance Index |
| ARVI | Atmospherically Resistant Vegetation Index |
| CHL-RED-EDGE | Chlorophyll Red-Edge |
| REPO | Red-Edge Position Index |
| EVI | Enhanced Vegetation Index |
| EVI2 | Enhanced Vegetation Index 2 |
| GNDVI | Green Normalized Difference Vegetation Index |
| MCRI | Modified Chlorophyll Absorption in Reflectance Index |
| MI | Moisture Index |
| NDMI | Normalized Difference Moisture Index |
| NDWI | Normalized Difference Water Index |
| NDMIMS | Normalized Difference Water Index for Moisture Stress |
| NDCI | Normalized Difference Chlorophyll Index |
| PSSRb1 | Pigment Specific Simple Ratio for Chlorophyll B |
| SAVI | Soil Adjusted Vegetation Index |
| SIPI | Structure Insensitive Pigment Index |
| PSRI | Plant Senescence Reflectance Index |
| NDYI | Normalized Difference Yellowness Index |

Appendix Table A2. Plant Traits Used in This Study

| Trait Acronym | Full Name |
| --- | --- |
| LAI | Leaf Area Index |
| $C_{ab}$ | Leaf Chlorophyll Content |
| CCC | Canopy Chlorophyll Content |
| FAPAR | Fraction of Absorbed Photosynthetically Active Radiation |
| FCOVER | Fraction of Green Vegetation Cover |